\documentclass[aps,prl,reprint,showpacs,floatfix,amsmath,amssymb,superscriptaddress]{revtex4-1}
\usepackage{graphicx}

\usepackage{amsmath}
\usepackage{physics}
\usepackage{bm}
\usepackage{here}
\begin{document}

\newcommand{\Yb}[1]{$^{#1}\text{Yb}$}
\newcommand{\state}[3]{$^{#1}#2_{#3}$}

\title{Antiferromagnetic Interorbital Spin-Exchange Interaction of \Yb{171}}
\author{Koki Ono}
\altaffiliation{Electronic address: koukiono3@yagura.scphys.kyoto-u.ac.jp}
\affiliation{Department of Physics, Graduate School of Science, Kyoto University, 606-8502, Japan}
\author{Jun Kobayashi}
\affiliation{PRESTO, Japan Science and Technology Agency, Kyoto 606-8502, Japan}
\author{Yoshiki Amano}
\affiliation{Department of Physics, Graduate School of Science, Kyoto University, 606-8502, Japan}
\author{Koji Sato}
\affiliation{Department of Physics, Graduate School of Science, Kyoto University, 606-8502, Japan}
\author{Yoshiro Takahashi}
\affiliation{Department of Physics, Graduate School of Science, Kyoto University, 606-8502, Japan}
\date{\today}
\begin{abstract}
We report on the investigation of the scattering properties between the ground state \state{1}{S}{0} and the metastable state \state{3}{P}{0} of the fermionic isotope of \Yb{171}. We measure the $s$-wave scattering lengths in the two-orbital collision channels as $a_{eg}^+=225(13)a_0$ and $a_{eg}^-=355(6)a_0$, using clock transition spectroscopy in a three-dimensional optical lattice. The results show that the interorbital spin-exchange interaction is antiferromagnetic, indicating that \Yb{171} is a promising isotope for the quantum simulation of the Kondo effect with the two-orbital system.  
\end{abstract}
\maketitle

\section{I. INTRODUCTION}
Ultracold atomic gases have been successfully used to study quantum many-body systems owing to a high degree of controllability \cite{Greiner2002}. Thus far, the single-band Hubbard model has been the main target of quantum simulations of condensed-matter systems using ultracold atoms, revealing a great deal of important physics \cite{Gross2017}.
However, real materials such as transition metal oxide exhibit  rich orbital degrees of freedom as well as spin, the description of which is beyond the single-band Hubbard model. In this respect, alkaline-earth-like atoms have received much attention in recent years as an important experimental platform for unique quantum simulations \cite{Taie2012}. One of the remarkable properties of two-electron atoms is the existence of metastable states \state{3}{P}{0} or \state{3}{P}{2} as well as the ground state \state{1}{S}{0}. Fermionic isotopes of alkaline-earth atoms in the ground state $\ket{g}=\ket{^1S_0}$ and in the metastable state $\ket{e}=\ket{^3P_0}$ trapped in an optical lattice can be described by the two-orbital SU($\mathcal{N}$) Hubbard Hamiltonian, including the spin-exchange interaction term between $\ket{g\uparrow}$ and $\ket{e\downarrow}$, where the arrows represent arbitrary components of the nuclear spin $I$ \cite{Gorshkov2010}. 

One of the most important problems in condensed matter physics, which highlights a relevant role of the orbital and spin degrees of freedom, is the Kondo effect \cite{Kondo1964}, in which 
 an impurity in one orbit forms a spin-singlet state with an electron in the conduction band in the other orbit, inducing a Fermi surface instability.
The Kondo effect manifests itself in the increase of the resistance at low temperature, in contrast to the monotonic decrease expected for non-interacting fermions.  
The competition between the magnetic correlation and localization effects is believed to induce rich quantum phases represented by a Doniach phase diagram \cite{Doniach1977}. 
Several  schemes for cold-atom quantum simulation of the Kondo effect have been proposed for alkali atoms, which require superlattice structures \cite{Paredes2005} or a population of excited
bands \cite{Duran2004}, and a confinement-induced $p$-wave resonance \cite{Nishida2013,Nishida2016}. 
However, so far there have been no reports on the experimental progress of these proposals.  
In contrast, the two-orbital system naturally realized in the two-electron atoms is a promising candidate for the quantum simulation of the Kondo effect \cite{Foss-Feig2010,Nakagawa2015,Zhang2016,Kanasz-Nagy2018a,Nakagawa2018}. 

One of the essential ingredients for the emergence of the Kondo effect is an antiferromagnetic coupling. The scattering properties between the \state{1}{S}{0} and \state{3}{P}{0} states in fermionic isotopes of \Yb{173}($I=5/2$) \cite{Scazza2015, Cappellini2014,Hofer2015,Pagano2015} and $^{87}$Sr($I=9/2$) \cite{Zhang2014} have been investigated previously, and the interorbital spin-exchange interactions were found to be ferromagnetic with $a_{eg}^+=1878(37)a_0$ and $a_{eg}^-=220(2)a_0$ for \Yb{173} and $a_{eg}^+=169(8)a_0$ and $a_{eg}^-=68(22)a_0$ for $^{87}$Sr, which are the $s$-wave scattering lengths in units of the Bohr radius $a_0$ in the nuclear spin-singlet state $\ket{eg^+}$ and triplet state $\ket{eg^-}$, respectively. In Ref.~\cite{Riegger2018}, a genuine scheme of tunable spin-exchange interaction of \Yb{173} using a confinement-induced resonance \cite{Olshanii1998} was successfully demonstrated and found to be consistent with the theoretical results in Ref.~\cite{Zhang2018}, but at the same time the particle loss from the trap was observed.  
\begin{figure*}
\centering
\includegraphics[width=1\linewidth]{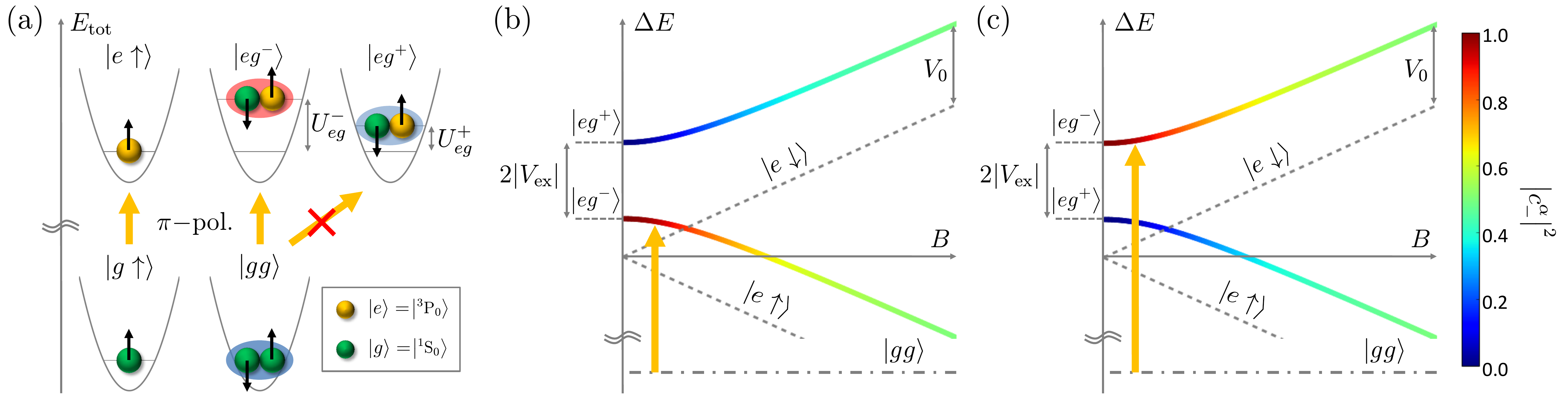}
\caption{(a) Singly and doubly occupied sites of \Yb{171}($I=1/2$) in an optical lattice at a zero magnetic field. A green (yellow) ball with a black arrow is the \state{1}{S}{0} (\state{3}{P}{0}) atom in a nuclear spin state $m_{F}=\pm1/2$. Yellow arrows indicate single-photon excitations to the \state{3}{P}{0} state using the $\pi$-polarized clock light. A blue (red) ellipse shows a nuclear spin singlet (triplet) state. Note that the interatomic interaction between  \Yb{171} atoms in the ground state is much smaller than that between $e$--$g$ atoms \cite{Kitagawa2008}. (b),(c) Illustrative plots of the magnetic field dependence of the eigenenergies for the singly occupied sites in the \state{3}{P}{0} state (dashed lines), the interorbital doubly occupied sites (solid curves with color gradient) and the doubly occupied sites in the ground state (dot-dashed lines) in the case of (b) ferromagnetic and (c) antiferromagnetic spin-exchange interaction. The line color shows the absolute square of the spin-triplet amplitude $\abs{c_-^\alpha}^2$ of the eigenstates $\ket{eg^\alpha}=c_+^\alpha(B)\ket{eg^+}+c_-^\alpha(B)\ket{eg^-}$, where $\alpha=\text{H}$ and $\text{L}$ represent the higher and lower branch, respectively. Yellow arrows indicate the optical coupling with the $\pi$-polarized light for the doubly occupied sites.}
\label{fig1}
\end{figure*}

In this work, we report on the measurement of the interorbital spin-exchange interaction of \Yb{171}($I=1/2$). Its scattering properties remain unexplored among the fermionic isotopes of the two-electron atoms cooled below the Fermi temperature $T_\text{F}$ whereas the $p$-wave scattering has been studied at a high temperature of $T\sim10\;\mu$K \cite{Ludlow2011,Lemke2011,Yanagimoto2018}. The clock transition spectroscopy is performed after loading the atoms into a three-dimensional (3D) optical lattice with a magic wavelength of 759~nm. We successfully measure the resonances from singly occupied sites and doubly occupied sites. Using systematic measurements of the resonances at various magnetic fields, we obtain $a_{eg}^+=225(13)a_0$ and $a_{eg}^-=355(6)a_0$. The results show that the spin-exchange interaction between the two-orbital states is antiferromagnetic $a_{eg}^+-a_{eg}^-=-131(19)a_0<0$. With lattice depths given with respect to \state{1}{S}{0} state atoms and $k_\text{B}$ as the Boltzmann constant, the spin-exchange interaction in a combined transverse optical lattice at the magic wavelength of depth $k_\text{B}\times 1$~$\mu$K and a longitudinal lattice formed by a laser at 655~nm of depth $k_\text{B}\times0.5$~$\mu$K is on the order of $k_\text{B}\times10$~nK. This is comparable to the currently
 achieved temperature of the atoms in the optical lattice. This work paves the way to the quantum simulation of the Kondo effect.
 
\section{II. INTERORBITAL SPIN-EXCHANGE INTERACTION}

We consider the scattering properties between the \state{1}{S}{0} and \state{3}{P}{0} atoms \cite{Scazza2015}. Atoms in the different orbitals with different nuclear spins collide via the following two anti-symmetric channels:
\begin{equation}
\ket{eg^\pm}=(\ket{eg}\pm\ket{ge})(\ket{\uparrow\downarrow}\mp\ket{\downarrow\uparrow})/2.
\label{eq1}
\end{equation}
Thus, the onsite Hamiltonian in an optical lattice can be diagonal in the nuclear spin singlet-triplet basis, and the interorbital Hubbard interaction is written as 
\begin{equation}
U_X=\frac{4\pi\hbar^2}{m}{a_X}\int\dd^3\vb*{r}\abs{w_g(\vb*{r})}^2\abs{w_e(\vb*{r})}^2,
\label{eq2}
\end{equation} 
where $a_X$ represents the $s$-wave scattering length associated with the scattering channel $X=eg^\pm$. Here $m$ is the atomic mass and $w_\alpha(\vb*{r})$ ($\alpha=e,g$) is the lowest-band Wannier function. Relevant transition channels in an optical lattice are illustrated in Fig.~\ref{fig1}(a). As the optical coupling with the $\pi$-polarized clock light is allowed only for the transition $\ket{gg}\to\ket{eg^-}$, the energy difference $U_{eg}^--U_{gg}$ can be directly obtained by measuring the frequency shift between the resonances from singly and doubly occupied sites in an optical lattice.

In a non zero magnetic field $B$, the Zeeman interaction mixes the nuclear spin-singlet with the spin-triplet states, and the onsite Hamiltonian $H_{eg}$ in the $\{\ket{eg^+},\ket{eg^-}\}$ basis is
\begin{equation}
H_{eg}=\mqty(U_{eg}^+ & \Delta(B) \\ \Delta(B) & U_{eg}^-),
\label{eq3}
\end{equation}
where $\Delta(B)=\delta g m_{F}\mu_B B$ is the differential Zeeman shift between the \state{1}{S}{0} and \state{3}{P}{0} states. Here $m_{F}$ denotes the nuclear spin projection along the magnetic field, $\mu_B$ is the Bohr magneton, and $\delta g=g_e-g_g$, where $g_g$ and $g_e$ represent the nuclear $g$-factors in the \state{1}{S}{0} ground and the \state{3}{P}{0} metastable states, respectively. The eigenenergies of the Hamiltonian (\ref{eq3}) are
\begin{equation}
E^\alpha(B)=V_{0}\pm\sqrt{V_{\text{ex}}^2+\Delta(B)^2},
\label{eq4}
\end{equation}
where $V_{0}=(U_{eg}^++U_{eg}^-)/2$ is the direct interaction and $V_{\text{ex}}=(U_{eg}^+-U_{eg}^-)/2$ is the interorbital nuclear spin-exchange interaction. Here $\alpha=\text{H}$ and $\text{L}$ correspond to the higher and lower branch, respectively. The sign of the spin-exchange interaction $V_{\text{ex}}$ is especially important because it characterizes the magnetism in the ground state of the two-orbital system. For $V_{\text{ex}}>0 \ (V_{\text{ex}}<0)$, a nuclear spin-triplet (spin-singlet) has the lowest energy indicating a ferromagnetic (an antiferromagnetic) spin-exchange interaction. Figures \ref{fig1}(b) and \ref{fig1}(c) show the two eigenenergies as a function of a magnetic field for \ref{fig1}(b) ferromagnetic and \ref{fig1}(c) antiferromagnetic interactions. In the antiferromagnetic case shown in Fig.~ \ref{fig1}(c), for example, the higher (lower) of the two colored branches connects to a spin-triplet (singlet) state, associated with a red (blue) point, at a zero magnetic field.  As the magnetic fields are increased, the branches asymptotically approach the superpositions of $\ket{eg^+}$ and $\ket{eg^-}$ states, associated with the green lines.  Thus, clock transition spectroscopic measurements in an optical lattice at various magnetic fields enable us to determine $V_0$ and $V_{\text{ex}}$ by fitting the resonance positions with Eq.~(\ref{eq4}).  
\section{III. METHODS}
Our experiment starts with the preparation of quantum degenerate gases of \Yb{171} by sympathetic evaporative cooling with \Yb{173} atoms in a crossed dipole trap \cite{Taie2010}. The number of atoms $N$ and the temperature $T$ in the trap are typically $N=1.0\times10^4$ and $T/T_{\text{F}}\sim 0.2$. After the evaporation, the remaining \Yb{173} atoms are removed by shining the resonance light on the \state{1}{S}{0}--\state{3}{P}{1} ($F=3/2$) transition at 556~nm (see Appendix A for the relevant energy diagram). The atoms are loaded into a 3D optical lattice with a magic wavelength of 759~nm in 200~ms, and the ground state atoms are excited to the \state{3}{P}{0} ($F=1/2$) state by the $\pi$-polarized clock light at 578~nm with a duration of 50~ms. The clock light is generated by sum-frequency generation in a periodically poled lithium niobate module using two pump lasers with the wavelengths of 1030 and 1319~nm.  The frequency stabilization of the clock laser is done by locking the laser to an ultra-low-expansion cavity, resulting in about 1 kHz linewidth with less than 1 kHz/h frequency drift. Then the remaining atoms in the ground state are removed from the optical lattices by irradiating a laser pulse of 1 ms resonant to the \state{1}{S}{0}--\state{1}{P}{1} transition at 399~nm. The blast pulse generally causes heating of the sample. However, heating of the \state{3}{P}{0} atoms is not detrimental to the spectroscopy measurement in which we apply the blast pulse only after the spectroscopy excitation pulse resonant to the \state{1}{S}{0}--\state{3}{P}{0} transition. The atoms in the \state{3}{P}{0} state are repumped to the \state{1}{S}{0} state via the \state{3}{S}{1} state by two laser pulses of 1 ms, whose wavelengths are 649~nm for the \state{3}{P}{0}--\state{3}{S}{1} transition and 770~nm for the \state{3}{P}{2}--\state{3}{S}{1} transition. Finally, the repumped atoms are captured by a magneto-optical trap using the strong \state{1}{S}{0}--\state{1}{P}{1} ($F=3/2$) transition with a magnetic gradient of 45 Gauss/cm, and the fluorescence from the trap is detected with an electron multiplying charge-coupled device camera, which enables a high detection sensitivity of  fewer than 100 atoms. Note that this scheme successfully reproduces the results of the previous experiments \cite{Scazza2015, Cappellini2014} for the measurement of the two-orbital interaction of \Yb{173}. 
\section{IV. RESULTS}
\begin{figure}
\begin{center}
\includegraphics[trim=0cm 2.5cm 0cm 2.5cm,clip,width=1\linewidth]{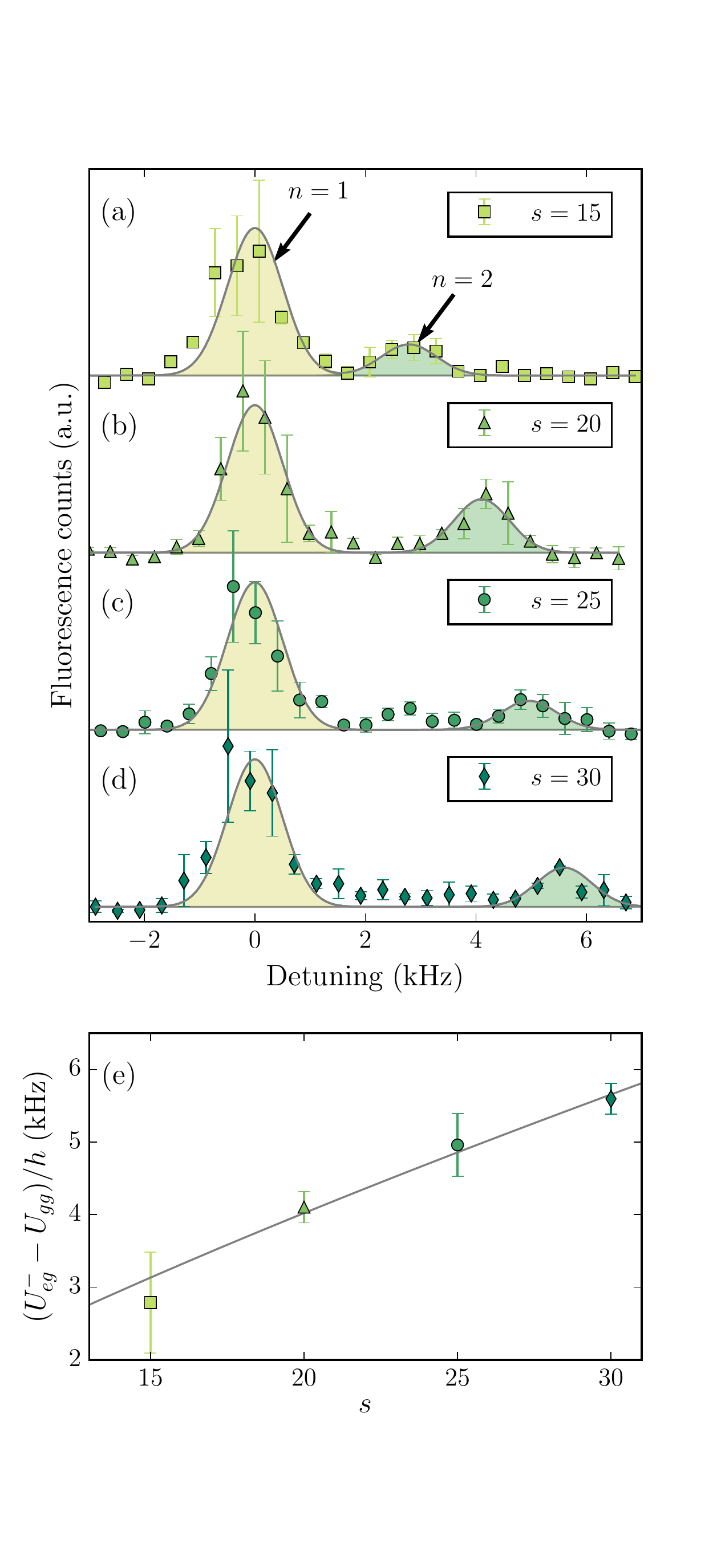}
\caption{Clock transition spectroscopy in a 3D optical lattice for different lattice depths: (a) $s=15$,; (b) $s=20$,; (c) $s=25$,; (d) $s=30$. The horizontal axis shows the detuning of the clock laser from the resonance of singly occupied sites. Two peaks labeled by $n$ correspond to resonances of singly and doubly occupied sites. Error bars show the standard deviations of the mean values obtained by averaging three measurements. (e) Interaction shift as a function of $s$. Error bars are 95 $\%$ confidence intervals of resonance position fits. The solid line represents fits to the data with Eq.~(\ref{eq2})}. 
\label{fig2}
\end{center}
\end{figure}

Figures \ref{fig2}(a)-\ref{fig2}(d) show the results of the clock transition spectroscopy in optical lattices at a zero magnetic field. Here $s$ denotes the lattice depth scaled by the recoil energy $E_\text{R}=\hbar^2/2m\lambda_\text{L}^2$, with $\lambda_\text{L}$ and $m$ being the lattice wavelength and the mass of \Yb{171}, respectively. We observe the resonances associated with singly occupied sites and doubly occupied sites. Here the data are plotted as a function of the detuning with respect to the resonance frequency for the singly occupied sites, which is easily identified from the expected magnetic field dependence explained in the following, and the robustness for the change of the atom density. The resonances observed at the higher-frequency side of resonance of the singly occupied sites are successfully identified as the resonances from the doubly occupied sites by confirming the disappearance of the peaks at low atom density above $T/T_\text{F}=0.3$ and also the fact that, as shown in Fig.~\ref{fig2}(e), the transition frequency depends on the lattice depth owing to the change of the on-site interaction according to Eq.~(\ref{eq2}), where the Wannier function changes upon the lattice depth in contrast to the resonances of singly occupied sites. The two-particle peaks associated with different Wannier orbitals other than those corresponding to the lowest band of $\ket{eg^-}$ branch were not clearly identified with the current signal-to-noise ratio in our experiment. Two-photon excitation is not expected to be observed since the excitation is chosen to be weak to suppress power broadening of the spectrum. From this measurement of  the interaction shifts $U_{eg}^--U_{gg}$ at various lattice depths, we obtain the $s$-wave scattering length in a spin-triplet state $a_{eg}^-=355(6)a_0$ through fitting the interaction shifts with the Hubbard interaction energy in Eq.~(\ref{eq2}). Here the error shows the standard deviation of the fits in Fig. \ref{fig2}(e). By also including higher band contributions the precision in the determination of the scattering lengths can be further improved. However, numerical calculations in Supplemental Material of Ref.~\cite{Cappellini2014} show that this correction is negligible compered with the error of the data. Previously, the $s$-wave scattering length of $a_{eg}^-=25a_0$ was inferred in Ref.~\cite{Lemke2011}, but we believe the uncertainty of the value should be large because the previous work was performed at the temperature of 10~$\mu$K, where $p$-wave collision is dominant, and a small contribution of the $s$-wave scattering is difficult to estimate accurately, as was mentioned in Ref.~\cite{Lemke2011}. 

\begin{figure}
\begin{center}
\includegraphics[trim=0cm 2cm 0cm 2cm,clip,width=1\linewidth]{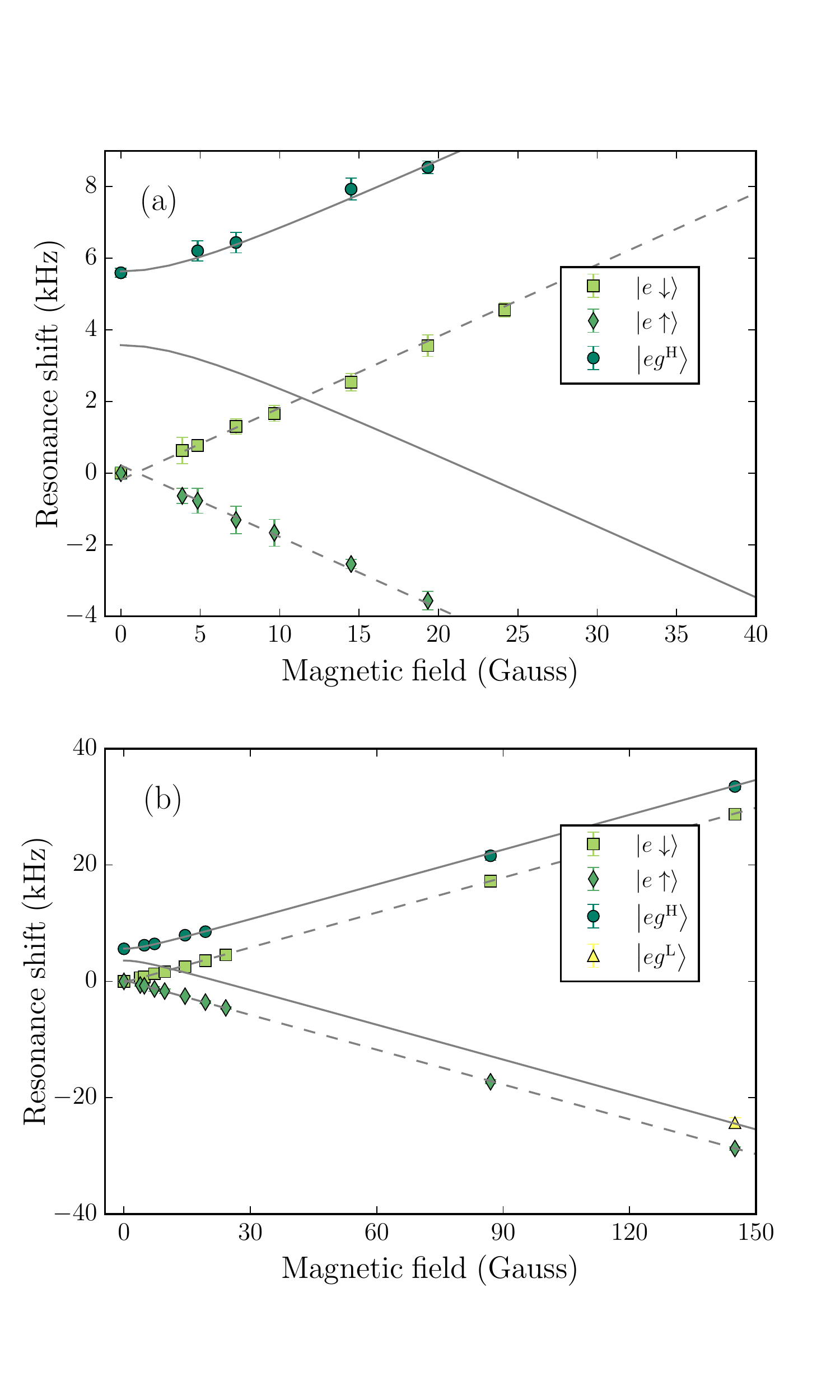}
\caption{(a) Magnetic field dependence of clock transition frequencies of \Yb{171} in 3D magic optical lattice. Squares mark transitions to $\ket{e\downarrow}$ and diamonds mark transitions to $\ket{e\uparrow}$. Circle (triangle, only at 145 Gauss) points indicate transitions to $\ket{eg^\text{H}}$ ($\ket{eg^\text{L}}$), which asymptotically connect to  $\ket{eg^-}$($\ket{eg^+}$) in a zero magnetic field. The lattice depth is $(s_x,s_y,s_z)=(30,30,30)$, where $s_i$ ($i=x,y,z$) denotes the lattice depth in the units of the recoil energy along the $i$ axis. Error bars are 95 $\%$ confidence intervals of resonance position fits. Solid lines represent the fits to the data with Eq.~(\ref{eq4}). (b) Resonance positions as a function of a magnetic field extending to higher magnetic fields.} 
\label{fig4}
\end{center}
\end{figure}

In addition, we measure the transition frequencies at various magnetic fields in a 3D optical lattice of $s=30$ as shown in Fig.~\ref{fig4}.  As depicted in Fig.~\ref{fig1}(b), the resonances from singly occupied sites are well fitted with the two linear lines as a function of a magnetic field, with a slope of $\Delta(B)/(Bh)=-200.0(6)$ Hz/Gauss, which is in good agreement with previous work in Refs.~\cite{Porsev2004, Lemke2009}. The corresponding spin state for each resonance is also confirmed by observing the nuclear spin distribution after the excitation using an optical Stern--Gerlach scheme \cite{Taie2010}. The important features of the excitation associated with doubly occupied sites in the case of $V_{\text{ex}} <0$, depicted in Fig.~\ref{fig1}(c), are that in a lower magnetic field, the two-particle state is excited to the higher branch, asymptotically connected to a nuclear spin-triplet state $\ket{eg^-}$ at a zero magnetic field, whereas it is excited to both the higher and lower branches at higher magnetic fields owing to mixing between the spin-singlet and spin-triplet states. The observed experimental data in Figs.~\ref{fig4}(a) and \ref{fig4}(b) clearly show these features expected for $V_{\text{ex}}<0$. Fitting the observed transition frequencies in Fig.~\ref{fig4} with the eigenenergies in Eq. (\ref{eq4}) yields $V_{\text{ex}}/h=-1.03(15)$ kHz and $V_{0}/h=4.56(13)$ kHz for the lattice depth of 30 $E_\text{R}$, from which we can obtain the $s$-wave scattering lengths for nuclear spin-singlet and spin-triplet states $a_{eg}^+=225(13)a_0$ and $a_{eg}^-=356(13)a_0$, respectively. The $a_{eg}^-$ obtained by the measurement in Fig.~\ref{fig4} is consistent with the result in Fig.~\ref{fig2} within the error bar. 
\section{V. DISCUSSION}
We discuss the experimental feasibilities for the antiferromagnetic Kondo lattice model (KLM) proposed in Ref.~\cite{Gorshkov2010}, where the \state{1}{S}{0} and \state{3}{P}{0} states of \Yb{171} atoms correspond to mobile particles and localized spins, respectively. The \state{3}{P}{0} atoms can be localized in a quasi-periodic optical lattice by additionally introducing an optical lattice with a different lattice constant such as 650 nm, which is deep only for the \state{3}{P}{0} atoms, or in a state-dependent optical lattice \cite{Riegger2018}. The phase diagram of the KLM, called the Doniach phase diagram \cite{Doniach1977}, is characterized by spin correlation between itinerant atoms and localized spins. In the strong coupling regime, a heavy-Fermi-liquid behavior is expected when the temperature is below the Kondo temperature. The Kondo temperature in the state-dependent optical lattice is estimated to be approximately 10~nK, which is comparable with experimentally achievable temperature in the optical lattice (see Appendix C for the calculation of the Kondo temperature). The two-orbital system using \Yb{171} atoms, therefore, is quite promising for realization of the quantum simulation of the Kondo effect. In addition, the negligible interaction between atoms in the ground state offers another advantage that the atoms in the ground state \state{1}{S}{0} is well described as a non interacting metallic state. This is ideal for the study of the Kondo effect because the origin of the suppression of quantum transport are well separated from the interaction effect as in the Mott insulating phase. Effective mass enhancement of the delocalized atoms will be probed using the dipole oscillation scheme proposed in Ref. \cite{Foss-Feig2010}. In the weak coupling regime, on the other hand, atoms in the ground state can mediate the Ruderman--Kittel--Kasuya--Yoshida (RKKY) interaction \cite{Ruderman1954} between atoms in the \state{3}{P}{0} state, with a characteristic energy of $k_\text{B}T_{\text{RKKY}}\sim V_{\text{ex}}^2/J_g$. The modulated spin-exchange interaction induced by the RKKY interaction will be observed using double-well potentials as proposed in Ref.~\cite{Gorshkov2010}. 

\section{VI. CONCLUSION}
In conclusion, the clock transition spectroscopy of a quantum gas of \Yb{171} atoms in a 3D optical lattice has been performed successfully. We have measured the $s$-wave scattering lengths in the two interorbital collision channels and have found that the spin-exchange interaction is antiferromagnetic. This work opens the possibility of the quantum simulation of the Kondo effect using alkaline-earth-like atoms. In addition, our result for \Yb{171} provides useful information for the determination of the mass-scaling properties of the \state{1}{S}{0}--\state{3}{P}{0} interorbital scattering lengths of Yb atoms, which will also be useful for a possible optical  lattice clock with weakly bound molecules \cite{Borkovski2018}.

\section{ACKNOWLEDGEMENTS}
We acknowledge valuable discussions with Peng Zhang, Ippei Danshita, Shimpei Goto, Masaya Nakagawa, and Yuto Ashida.
This work was supported by Grants-in-Aid for
Scientific Research from MEXT/JSPS KAKENHI (No.~25220711, No.~26610121, No.~17H06138, No.~18H05405, 
and No.~18H05228), the Impulsing Paradigm Change
through Disruptive Technologies (ImPACT) program, and
JST CREST(No.\ JPMJCR1673).

\section{APPENDIX A: ENERGY LEVELS OF YTTERBIUM ATOM}
\begin{figure}[H]
\begin{center}
\includegraphics[trim=0cm 1cm 0cm 1.5cm,clip,width=0.8\linewidth]{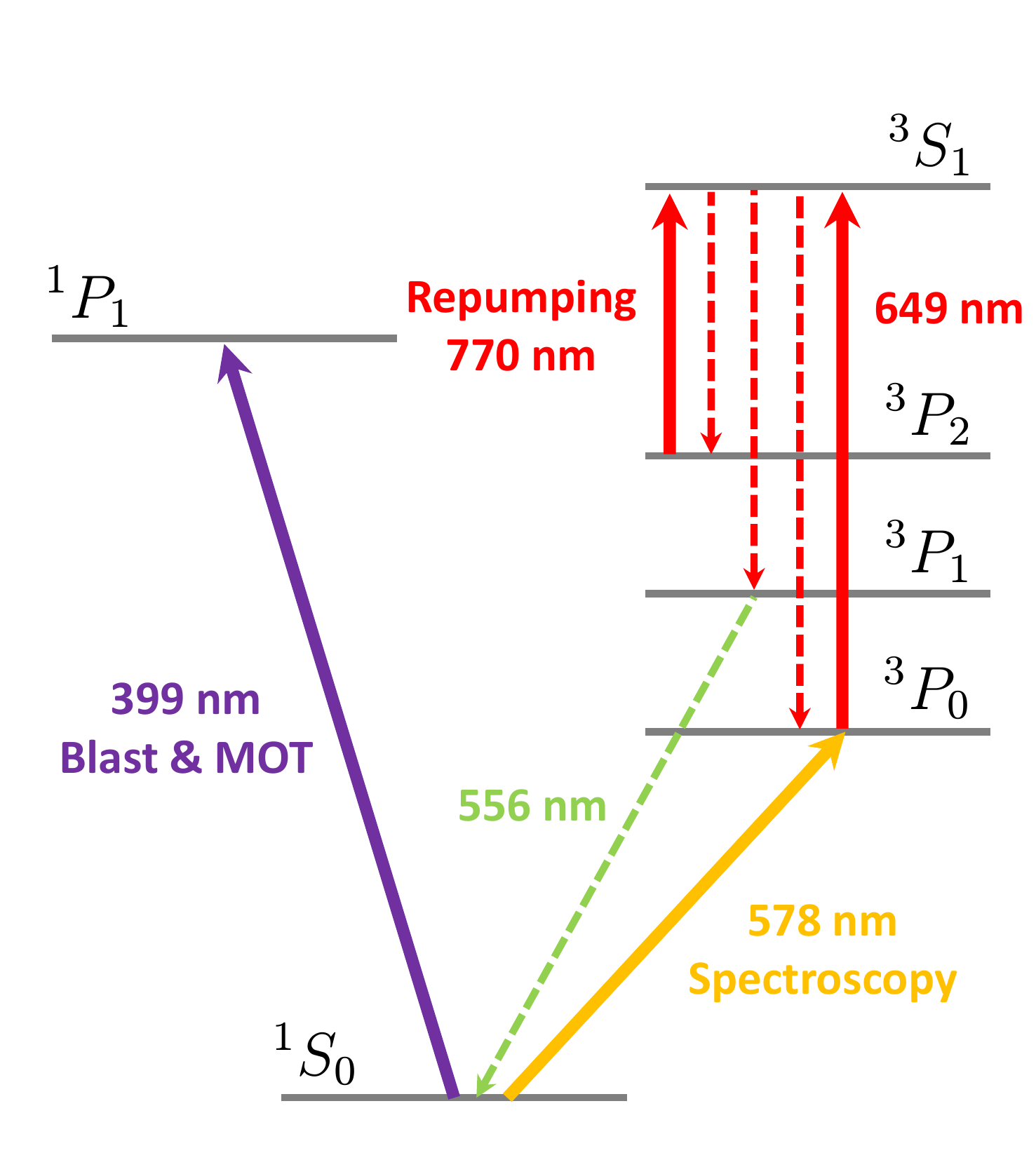}
\caption{Relevant energy levels for the clock transition spectroscopy. After the excitation to the \state{3}{P}{0} state, associated with a yellow arrow, remaining atoms in the ground state are blasted with a strong \state{1}{S}{0}$\to$\state{1}{P}{1} transition, corresponding to a purple arrow. Then the \state{3}{P}{0} atoms are repumped into the ground state via a \state{3}{S}{1}$\to$\state{3}{P}{1}$\to$\state{1}{S}{0} process using simultaneous applications of \state{3}{P}{0}$\to$\state{3}{S}{1} and  \state{3}{P}{2}$\to$\state{3}{S}{1} resonant light beams, associated with solid red arrows. Finally, the repumped atoms are captured with a magneto-optical trap using the \state{1}{S}{0}$\to$\state{1}{P}{1} transition. Dashed arrows represent relevant spontaneous decays.} 
\label{figs1}
\end{center}
\end{figure}

Figure \ref{figs1} shows the optical transitions related to the detection of the \state{3}{P}{0} atoms. Note that the branching ratios of the \state{3}{S}{1}$\to$\state{3}{P}{0} and \state{3}{S}{1}$\to$\state{3}{P}{2} transitions are  0.36 and 0.51, respectively \cite{Porsev1999}.
\section{APPENDIX B: LOCALIZATION OF ATOMS IN THE \state{3}{P}{0} STATE}
\begin{figure}
\begin{center}
\includegraphics[trim=0cm 0cm 0cm 0cm,clip,width=0.9\linewidth]{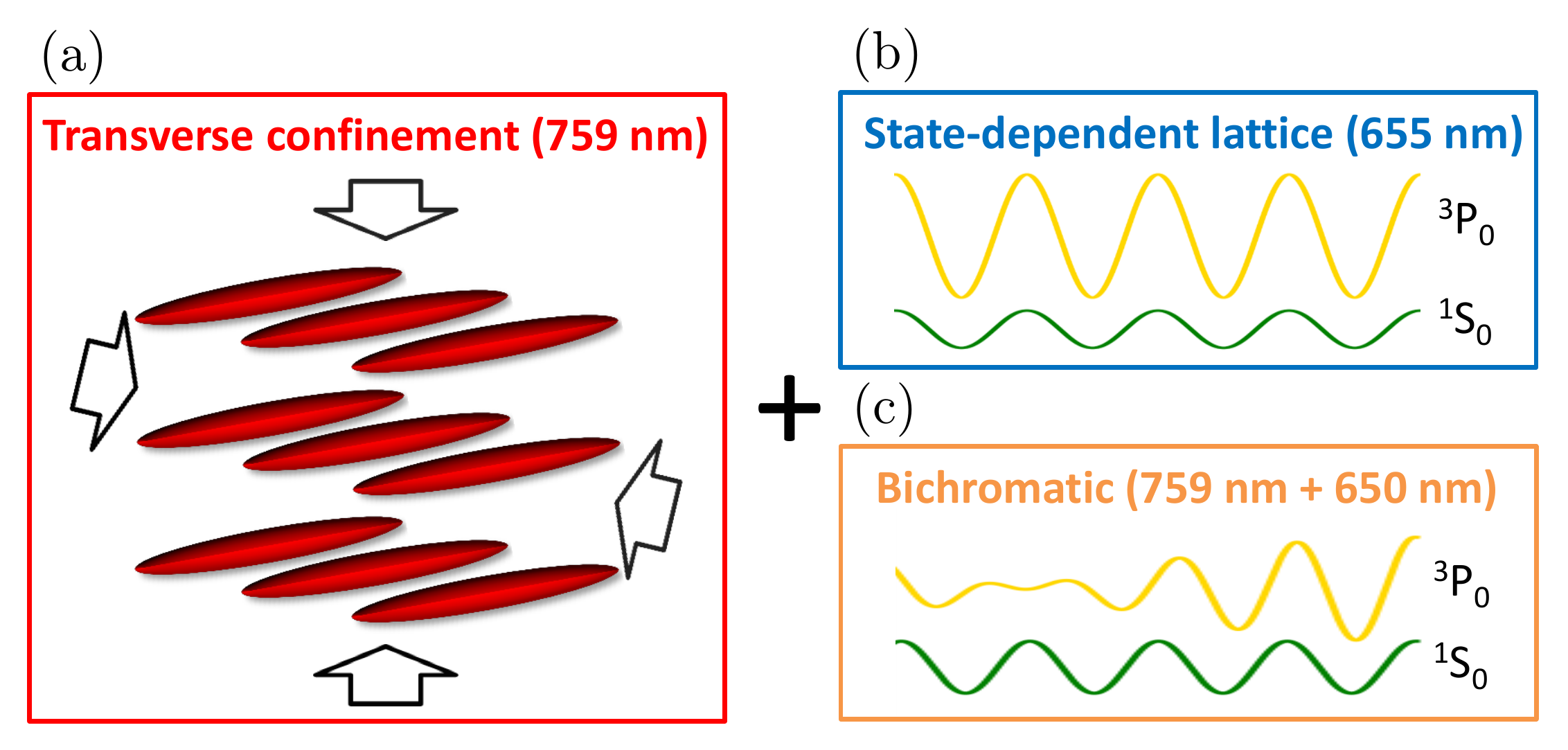}
\caption{Schematic diagram of the lattice geometry for the Kondo system. (a) Strong transverse potential with the wavelength of 759 nm to achieve a large value of $\abs{V_{\text{ex}}}$. (b) State-dependent lattice with a wavelength of 655 nm to localize the \state{3}{P}{0} atoms by the Anderson localization. (c) Bichromatic lattice using two beams with the wavelengths of 759 and 650 nm.} 
\label{figs2}
\end{center}
\end{figure}

 Figure \ref{figs2} shows a schematic diagram of the proposed lattice geometries for probing the Kondo effect. Strong transverse confinement is realized with a deep two-dimensional lattice with the magic wavelength of 759 nm as shown in Fig.~\ref{figs2}(a). We consider two kinds of longitudinal lattice potentials. Figure \ref{figs2}(b) shows the state-dependent potential with a wavelength of 655 nm, where the polarizability of the \state{3}{P}{0} state is 11 times larger than that of the \state{1}{S}{0} state. Figure \ref{figs2}(c) shows the bichromatic potential, which consists of a primary lattice with the magic wavelength and a secondary lattice with a wavelength of 650 nm, which gives much larger light shift to the \state{3}{P}{0} atom than the \state{1}{S}{0} atom. In this system, only the atoms in the \state{3}{P}{0} state experience the incommensurate potential,  resulting in the Anderson localization \cite{Roati2008}.
\section{APPENDIX C: CALCULATION OF KONDO TEMPERATURE}
\begin{figure}
\begin{center}
\includegraphics[trim=0cm 0cm 0cm 0.5cm,clip,width=0.9\linewidth]{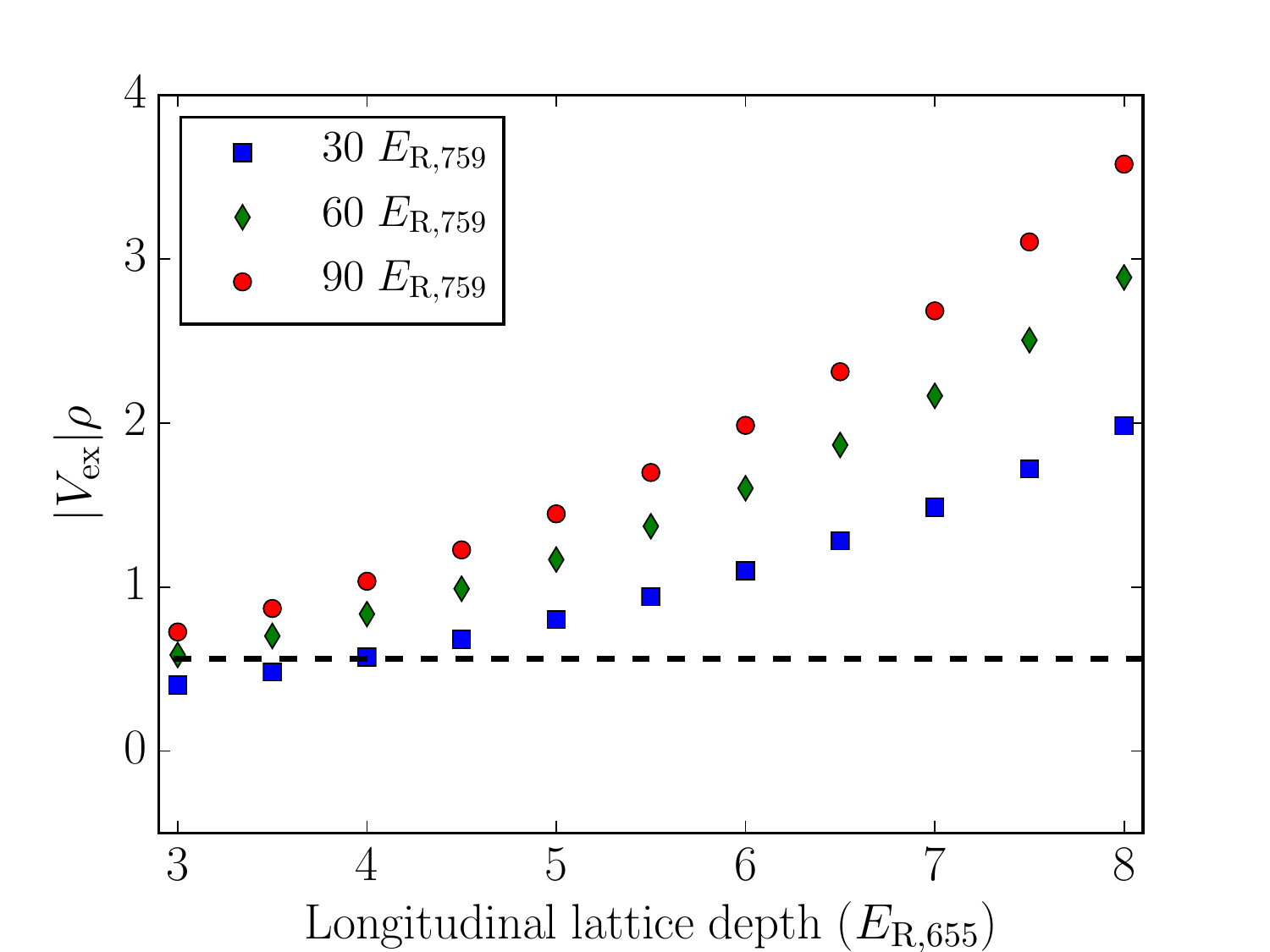}
\caption{Calculation of the dimensionless spin-exchange interaction of \Yb{171} in the state-dependent optical lattice ($\lambda= 655$ nm) to localize \state{3}{P}{0} atoms. A horizontal axis represents the longitudinal lattice depth for the \state{1}{S}{0} atom in units of the recoil energy $E_\text{R,655}=h^2/(2m\lambda^2)$. Circle, diamond, and square points indicate the lattice depths for the transverse confinement of 30, 60, and 90 $E_{\text{R,759}}$, respectively. Note that the expression for the Kondo temperature (\ref{S1}) is reliable below 0.6 indicated by a dashed line.}
\label{figs3}
\end{center}
\end{figure}

The Anderson scaling method shows that a renormalized spin-exchange interaction is enhanced with decreasing temperature and diverse at the Kondo temperature \cite{Hewson1997}, given by
\begin{equation}
T_\text{K}=D\sqrt{2\abs{V_\text{ex}}\rho}\exp\left(-\frac{1}{2\abs{V_\text{ex}}\rho}\right),
\label{S1}
\end{equation}
where $D$ and $\rho$ represent the band width and the density of states at the Fermi energy, respectively. This method examines how the T-matrix including scattering information between the conduction electrons and a magnetic impurity is changed when the high-energy electrons in the edge of the band are integrated out. The expression for the Kondo temperature \ref{S1} is valid in the weak coupling regime where higher-order terms $O((\abs{V_\text{ex}}\rho)^4)$ are negligible in the perturbative renormalization group approach for the Kondo model. In estimating the Kondo temperature in the one-dimensional configuration of the two orbital system, we assume $D=2J_g$ and $\rho=1/(2\pi J_g)$, where $J_g$ represents the tunneling energy of the atom in the \state{1}{S}{0} state. Figure \ref{figs3} shows the calculation of the dimensionless spin-exchange interaction $\abs{V_\text{ex}}\rho$ as a function of the longitudinal lattice depth for several transverse confinements in the case of the state-dependent lattice. Note that the expression for the Kondo temperature (\ref{S1}) is reliable for $\abs{V_\text{ex}}\rho$ below 0.6 indicated by a dashed line. For these values, the Kondo temperature is estimated to be about 10 nK, and the Kondo effect could emerge at an experimentally achievable temperature in an optical lattice. 

\section{APPENDIX D: EVALUATION OF LOCALIZATION OF \state{3}{P}{0} ATOM IN BICHROMATIC LATTICE}
\begin{figure}
\begin{center}
\includegraphics[trim=0cm 0cm 1.5cm 0cm,clip,width=0.9\linewidth]{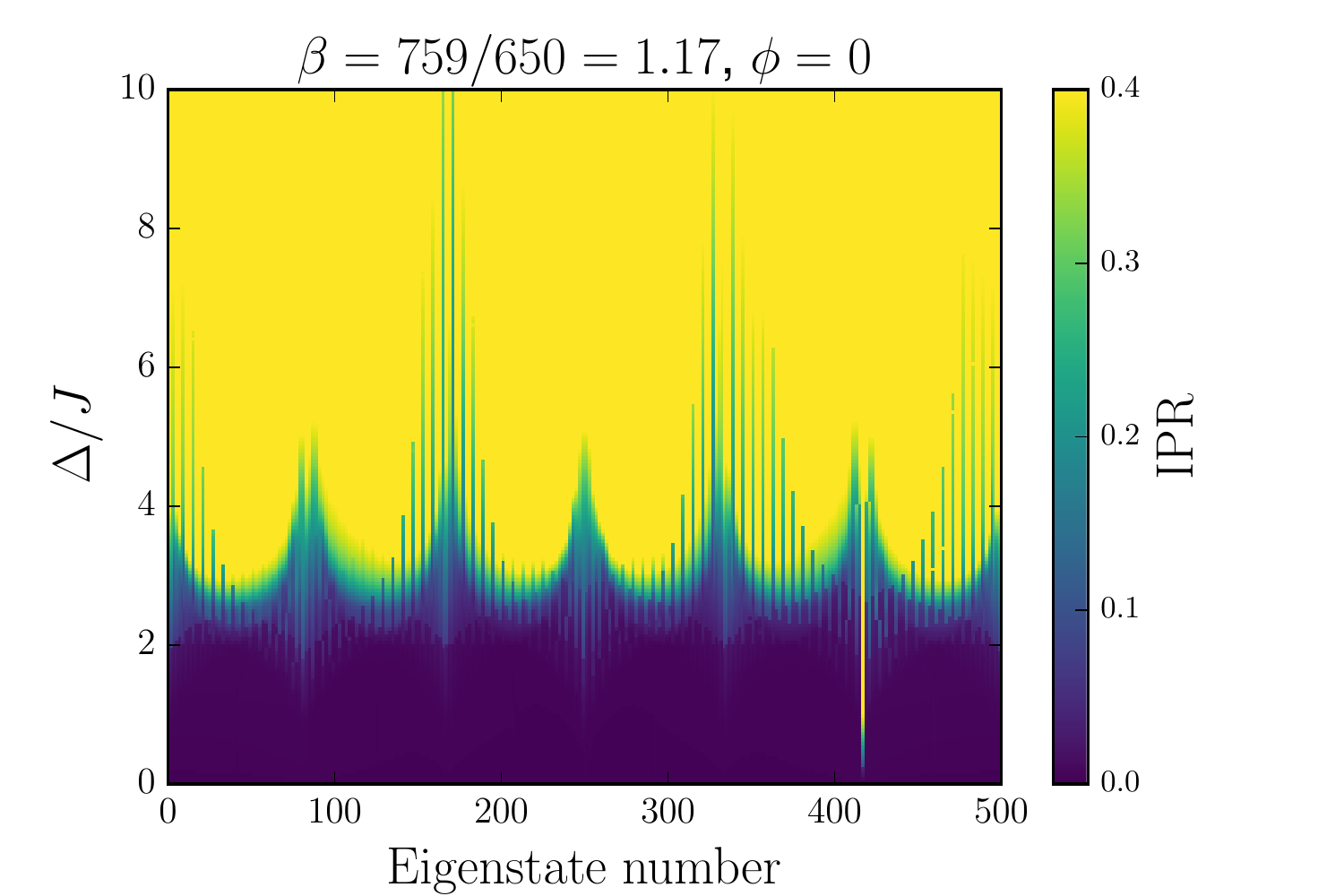}
\caption{Numerical calculation of the IPR in the bichromatic  lattice with system size of 500 sites. A vertical axis shows the quasi-periodic disorder strength $\Delta$ scaled by the hopping $J$. A horizontal axis indicates a label number assigned to an eigenstate. The relative phase
 $\phi$ is set to $0$.} 
\label{figs4}
\end{center}
\end{figure}

We consider a single-particle Hamiltonian in the bichromatic potential, given by
\begin{equation}
H=\frac{p^2}{2m}+s_1E_{\text{R},1}\sin^2(k_1x)+s_2E_{\text{R},2}\sin^2(k_2x+\phi),
\label{S2}
\end{equation}
where the indexes $i=1$ and $2$ correspond to the primary lattice and the secondary lattice, respectively. Here $k_i=2\pi/\lambda_i$ and $\phi$ are the wavelength number of the lattice and an arbitrary phase, respectively. In the tight-binding limit, the Hamiltonian in Eq.~(\ref{S2}) can be mapped to the Aubry--Andr{\'e} model \cite{Aubry1980}, defined by
\begin{equation}
H= -J\sum_j(c^\dagger_{j+1}c_j
+\text{H.c.})
+\Delta\sum_j\cos(2\pi\beta j+\phi)c^\dagger_j c_j,
\label{S3}
\end{equation}
with $\beta=\lambda_1/\lambda_2$ and $\Delta=s_2E_{\text{R},2}/2$. Here $c_j$ ($c_j^\dagger$) is the annihilation (creation) operator on the $j$-th site of the primary lattice with the wavelength $\lambda_1$, and $J$ represents the hopping energy. In order to evaluate the localization of the \state{3}{P}{0} atom in the quasi periodic potential, we introduce the inverse participation ratio (IPR) \cite{Modugno2009} $\sum_i\abs{\bra{w_i}\ket{\psi}}^4$, 
which measures the overlap between the eigenstate of the Hamiltonian $\ket{\psi}$ and the Wannier state of the $i$-th site $\ket{w_i}$. When a particle is maximally localized, the IPR is unity. Figure \ref{figs4} shows the numerical calculation of the IPR in the bichromatic lattice, as depicted in Fig.~ \ref{figs2}(c),  which consists of the primary lattice with $\lambda_1=759$ nm and $s_1=5$ and the secondary lattice with $\lambda_2=650$ nm. The result shows that almost all of the eigenstates are localized above $\Delta/J\sim3$, corresponding to $s_2=0.3$.
\bibliography{reference}
\end{document}